# 18. ATTRIBUTION OF EXTREME RAINFALL IN SOUTHEAST CHINA DURING MAY 2015

Claire Burke, Peter Stott, Ying Sun, and Andrew Ciavarella

*Anthropogenic climate change increased the probability that a short-duration, intense rainfall event would occur in parts of southeast China. This type of event occurred in May 2015, causing serious flooding.*

*Introduction.* The prerainy season in southern China usually starts in April with a rainbelt forming along the Indochina Peninsula. The rainbelt moves northward across eastern China throughout the rainy season, which generally lasts from May to August (Ding and Chan 2005). In 2015, the prerainy season began in early May, about a month later than normal. After the rainbelt was established, the rainfall was exceptionally heavy with the total precipitation in some southern provinces more than 50% greater than the 1971–2000 average (CMA 2016). The rain fell in a series of heavy storms, causing severe flooding in many cities with impacts that included loss of life.

We examine the change in the character of rainfall during May in terms of the number of consecutive days of rain, total rainfall over a period of $n$ days ($n$-day totals), and rainfall intensity. Using these metrics, we estimate the regional change in probability of extreme precipitation due to anthropogenic climate forcing.

*Data.* We use model data from a pair of multidecadal ensemble experiments using the latest Met Office HadGEM3-A-based attribution system. Each ensemble comprises 15 members spanning the period 1960–2013, one set with both anthropogenic and natural forcings (ALL) and the other with natural only (NAT). The system is an upgrade to that used in a number of previous studies described by Christidis et al. (2013) to higher resolution (N216 L85, 0.56° x 0.83° horizontally), the latest operational dynamical core (ENDGame; Wood and Stainforth 2010), and land surface model (JULES; Best et al. 2011), as well as an updated forcings set consistent with the CMIP5 generation (Jones et al. 2011). Members differ from one another solely through the stochastic physics and share atmospheric initialization from ERA–40 at 0000 UTC 1 December 1959.

We use observed daily precipitation data for 1961–2015 provided by the Climate Data Center of China National Meteorological Information Center (NMIC). This dataset uses quality controlled data from 2419 stations and is the best daily dataset available for climate study in China. Yang and Li (2014) show that most of the daily precipitation series are homogeneous and lack pronounced discontinuities resulting from instrumental changes or station relocation.

*Methods.* We divide the flood-affected area of southeast China into 12 regions of 3° × 3° areas with spatially coherent rainfall patterns and variability. Region locations and corresponding time series of total monthly rainfall for May are shown in Fig. 18.1. Monthly totals show no clear trends outside of interannual variability (see Supplemental Material for linear fits). There is also no clear separation of expected changes in monthly rainfall under anthropogenic and natural forcings. Given the large interannual variability of monthly totals and the nature of the floods in May 2015 being related to several large daily rainfall totals, we instead look at changes in intensity and duration of rainfall events.

For the month of May, we count the number of consecutive wet days (daily total rainfall >= 1 mm) (n_days) and record the total rainfall during each $n$-day event (n_day_tot). We then calculate the mean intensity of rainfall in mm day$^{-1}$ for each event (intens). Using this metric, any given month can have several rainfall events.

For each 3° × 3° region in southeast China, the mean of the observed station data and the mean of the grid cells for the model data are calculated for each day. This daily area-mean for each region is then used to calculate n_days, n_day_tot, and intens. We use all the available stations in each 3° × 3° grid box, generally 40–60 stations, though the number of stations varies from year to year.

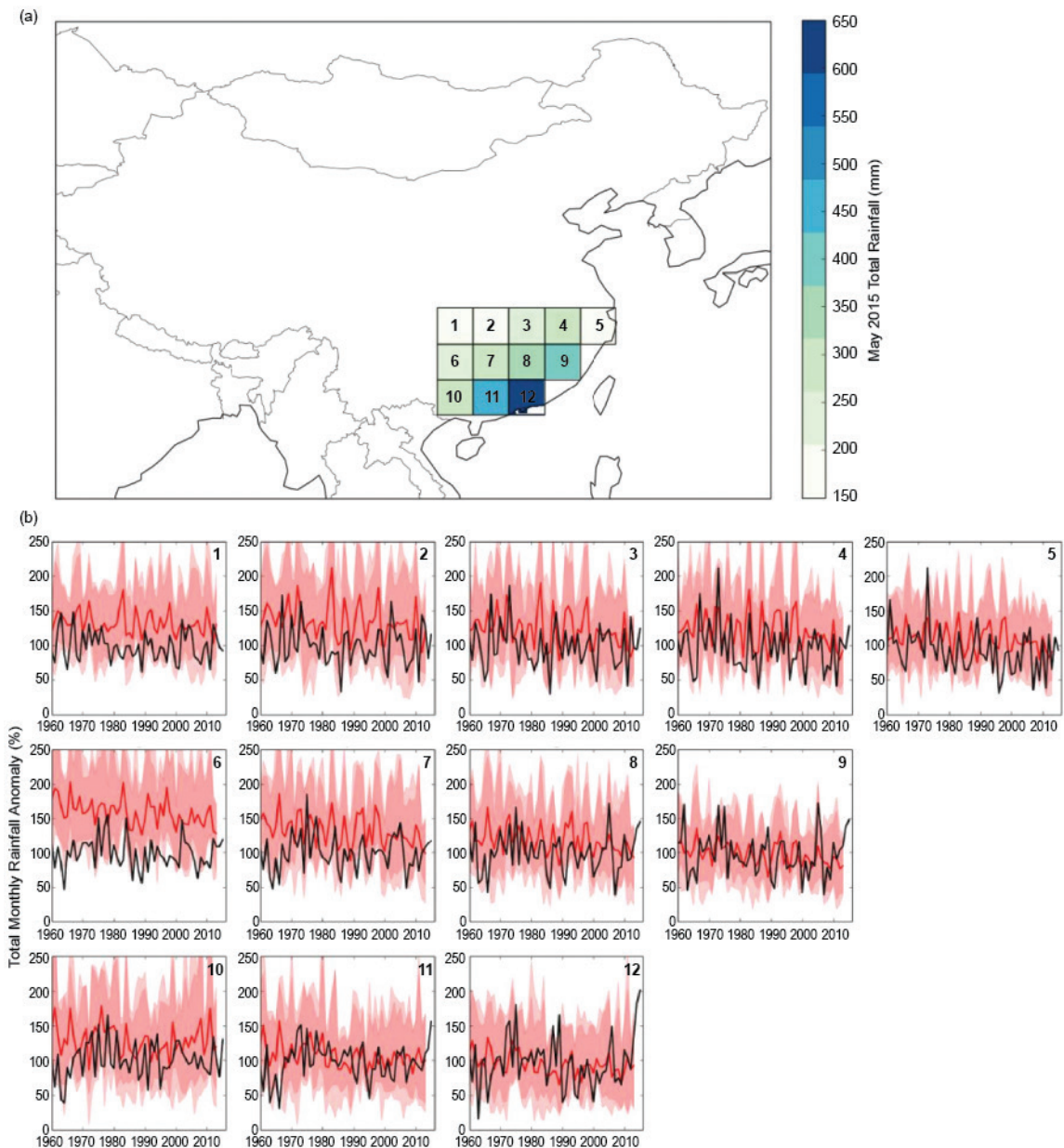

FIG. 18.1. (a) Regions of China examined in this study including total rainfall for May 2015. (b) Monthly total precipitation anomaly time series for regions 1–12: observed totals (black lines), HadGEM3-A model mean (thick red lines), model ensemble 5th–95th percentile (dark red shading), and total model ensemble range (light red shading). Anomalies for both observations and models are with respect to 1961–90 climatology.

*Model verification.* We assess the capability of the models to represent daily rainfall characteristics and extremes by comparison of time series and rainfall intensity distributions between the model ensemble and the observations.

We plot observed and model time series total monthly rainfall for May, total number of rainy days, and maximum intens for each of 12 regions in Fig. 18.1 and Supplemental Fig. S18.1. The observed interannual variability of the time series clearly falls within the ensemble variability of HadGEM3-A, with the exception of region 6, which we exclude from further analysis. Linear fits to the observed and modeled time series produce similar results albeit with low-fit significance (see Supplemental Material).

Supplemental Fig. S18.1 plots daily rainfall intensity distribution for the regions of southeast China. We perform a Kolmogrov–Smirnoff test (KS) for each region to determine how well the observed distribution of daily rainfall is reproduced by the ensemble mean—the results of this are indicated in the supplement. Nine of the 12 regions showed more than 95% likelihood of the observed and modeled distributions being drawn from the same population; an additional two regions showed 86%. Region 10 showed only 72% likelihood and was excluded from further analysis.

The model can reproduce the mean and extremes of precipitation totals, intensities, number of rainy days in a month, and numbers of consecutive days of rain sufficiently well for the attribution study intended here.

*Results.* We select the top 10% of $n\_day\_tot$ rainfall to be defined as extreme events and examine the change in intens and $n\_days$ over which this rain fell. The values for observed $n\_days$, $n\_day\_tot$, and intens are shown in Supplemental Fig. S18.2 for all years including 2015. Following Christidis and Stott (2015), we take the most recent 15 years of model data (1999–2013) as representative of current climatology to produce probability distribution functions (PDF) for intens and $n\_days$. The PDF is calculated by fitting a gamma function to the normalized model histogram of the rainfall metric examined; we test the appropriateness of this fit in the supplement. PDFs allow the calculation of fraction of attributable risk (FAR; Allen 2003), defined as FAR = 1 − (P(NAT) / P(ALL)) for individual regions.

We calculate FAR for intensity of events greater than the May 2015 observed maximum, providing an estimate of the extent to which human influence has increased the risk of high-total-rainfall events with daily intensity as high as observed in May 2015. We also calculate FAR for the number of consecutive days of rain less than the maximum value observed for May 2015, providing an estimate of the extent to which human influence has increased the risk of having high-total-rainfall events with duration as short as in May 2015.

To calculate the error on FAR, we bootstrap resample (with replacement) the top 10% of $n\_day\_tot$ for the all-forcings and natural-forcings models then refit the PDFs 1000 times for both intens and $n\_days$. The standard deviation of the FAR from the 1000 bootstrapped samples gives the error on FAR.

We exclude regions 6 and 10 from analysis (see model verification) and report results for the remaining 10 regions. We find positive FAR in 4 out of 10 and 9 out of 10 regions for intens and $n\_days$, respectively. The four regions for which intens showed positive FAR—7, 8, 11, and 12—are all spatially adjacent to each other; these and all the coastal regions show positive FAR for $n\_days$. The spatial contiguity of regions with positive FAR makes it more likely that these results are physically caused rather than a statistical fluke.

Three regions show positive FAR at 2σ in one metric and 1σ in the other; we focus our analysis on those. We find positive FAR at 2σ confidence for increase in intens and 1σ confidence for decrease in $n\_days$ in regions 7 and 12. We find positive FAR at 2σ confidence for decrease in $n\_days$ and 1σ confidence for increase in intens for region 11. We also find positive FAR for regions 8 and 9 at 1σ confidence for $n\_days$. We present results for regions 7, 11, and 12 in Fig. 18.2. For the top 10% of $n\_day\_tot$, we find anthropogenic climate change has increased the likelihood of intense rainfall, greater than or equal to that observed in May 2015, by 64% ± 17%, 23% ± 12%, and 66% ± 19% for regions 7, 11, and 12, respectively. For the same regions, we find anthropogenic influence increases likelihood by 12% ± 11%, 39% ± 14%, and 23% ± 14%, respectively, of a decrease in the number of consecutive days over which the rain fell with respect to the maximum number of consecutive days observed in May 2015.

Some studies show that the Pacific sea surface temperature anomaly (SSTA) is an important factor affecting the early rainy-season precipitation in southern China (e.g., Qiang and Yang 2013). We tested for the effect of El Niño on our results and find no obvious correlation between ENSO−3.4 index and any of the three indices we examine from May 2015.

*Conclusion.* During May 2015, large daily rainfall totals were recorded over much of southeast China. While no clear trends are seen in the monthly total rain for this region, we find that the character of the rainfall has changed, such that the same total amount of rain falls in shorter more intense storms. We have shown that for the month of May, anthropogenically forced climate change has increased the probability of this kind of intense, short-duration rainfall (as occurred in 2015) for some regions of southeast China. In the future, we might expect more occurrences of short, intense rainfall events in these regions, increasing the likelihood of flooding.

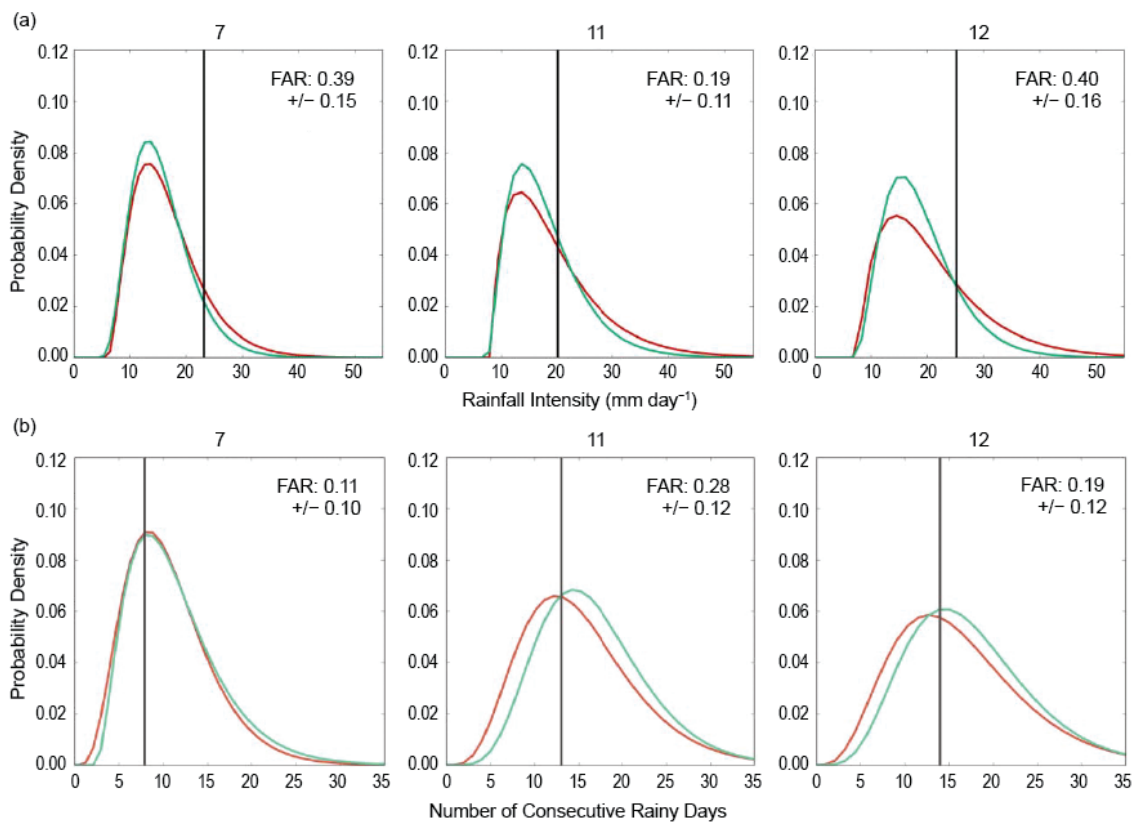

**Fig. 18.2.** (a) Probability distribution functions (PDFs) for the intensity of rainfall and (b) number of consecutive days of rain, where only the top 10% of rainfall total events (n_day_tot) defined by the observed 1961–90 climatology are selected. Red lines correspond to HadGEM3-A all forcings models (ALL); green is natural forcings only models (NAT); and black lines show the maximum value for May 2015. FARs (FAR = 1 − (P(NAT) / P(ALL))) are indicated for each region. (a) FAR for intensity greater than the 2015 maximum value and (b) FAR for number of consecutive days of rain less than the 2015 maximum value. Only regions 7, 11, and 12 are shown.

**ACKNOWLEDGEMENTS.** This work and two of its contributors (Burke and Ciavarella) were supported by the UK-China Research and Innovation Partnership Fund through the Met Office Climate Science for Service Partnership (CSSP) China as part of the Newton Fund. This work was supported by the Joint DECC/Defra Met Office Hadley Centre Climate Programme (GA01101). Ying Sun is supported by China funding grants GYHY201406020 and 2012CB417205.